\documentclass[aps,preprint,tightenlines,nofootinbib,amssymb,superscriptaddress]{revtex4}
\usepackage{amssymb}
\usepackage{mathrsfs}
\usepackage{slashed}
\usepackage{graphicx,color}
\usepackage{amsmath}
\allowdisplaybreaks

\begin{document}
\title{A Naturally Light Sterile Neutrino in an Asymmetric Dark Matter Model }
\author{Yongchao Zhang}
\email{yongchao@umd.edu}
\affiliation{Center for High Energy Physics, Peking University, Beijing, 100871, P. R. China}
\affiliation{Maryland Center for Fundamental Physics and Department of Physics, University of Maryland, College Park, Maryland 20742, USA}
\author{Xiangdong Ji}
\email{xji@umd.edu}
\affiliation{Center for High Energy Physics, Peking University, Beijing, 100871, P. R. China}
\affiliation{Maryland Center for Fundamental Physics and Department of Physics, University of Maryland, College Park, Maryland 20742, USA}
\affiliation{INPAC, Department of Physics and Astronomy, Shanghai Jiao Tong University, Shanghai, 200240, P. R. China}
\author{Rabindra N. Mohapatra}
\email{rmohapat@umd.edu}
\affiliation{Maryland Center for Fundamental Physics and Department of Physics, University of Maryland, College Park, Maryland 20742, USA}
\date{\today}
\hfill{UMD-PP-013-010}

\begin{abstract}
  A recently proposed asymmetric mirror dark matter model where  the mirror sector is connected with the visible one by a right-handed neutrino portal, is shown to lead naturally to a 3+1 active-sterile neutrino spectrum, if the portal consists only of two right-handed neutrinos.  At tree level the model has four massless neutrino states, three (predominately) active and one sterile. The active neutrinos pick up tiny masses via the minimal radiative inverse seesaw mechanism at one-loop level. The loop effects also generate the large solar and atmospheric mixings, as well as the observed reactor mixing for certain ranges of parameters of the model. The dominant contribution to the sterile neutrino mass ($\sim$ eV) arises from the gravitationally induced dimension-5 operators. Generating active-sterile mixing requires a two Higgs doublet extension of the standard model and a small mixing between the ordinary and mirror Higgs fields, which occurs naturally in mirror models.
  \end{abstract}
\maketitle
\tableofcontents

\newpage
\section{Introduction}
The fact that the three known neutrino species $(\nu_e, \nu_\mu, \nu_\tau)$ have masses and mix among themselves, has now been conclusively established by many neutrino oscillation searches, that involved heroic experimental efforts during the past two decades. These three neutrinos, known as active neutrinos, participate in weak interactions with full strength and are part of the standard model (SM), even though understanding their masses and mixings requires new physics beyond SM and is one of the most active areas of physics research right now. During the past several years, however, there have appeared indications from several experiments that there may be additional neutrinos, called sterile neutrinos (denoted by $\nu_s$) which do not directly participate in the usual weak interactions. They, however, mix with active neutrinos to produce new effects in various experiments. These experiments are: the LSND experiment~\cite{LSND}, the MiniBooNE neutrino and anti-neutrino experiments~\cite{MiniBooNE}, two classes of experiments, one indicating a deficit in the reactor neutrino flux compared to theoretical calculations (known as the reactor anomaly)~\cite{reactor-anomaly}, and another with similar deficit in the Gallium neutrino spectrum~\cite{Gallium-anomaly}. The new  sterile neutrinos, which could explain some or all of these anomalies, must have masses $\sim$ eV in order to fit the observations~\cite{sn-fit-Conrad,sn-fit-Giunti1,sn-fit-Giunti2,sn-fit-Kopp}. It is not clear how many such extra neutrino species are required by data. Furthermore, the compatibility of short baseline (SBL) experiments~\cite{sn-fit-planck} with the precision data from the Planck satellite~\cite{Planck}, which indicates the presence of dark radiation, is still a matter of intense discussion, but there is the possibility that this dark radiation could also be a manifestation of sterile neutrinos. In view of all these indications for a light sterile neutrino, there has been vigorous renewed interest in this field~\cite{abaz} with many proposed experiments to test these indications. On the theoretical side, there are many efforts to study the implications of light sterile neutrinos for physics beyond the standard model (BSM). The current work is an effort in this direction and what we find interesting is that our work is a direct out growth of an effort to understand the nature of dark matter and is not a BSM scenario just to explain the sterile neutrinos.

An obvious challenge that theoretical attempts to understand sterile neutrinos must overcome is that since the sterile neutrino is a singlet under the SM gauge group, there is no reason for it to be as light as an eV as required to fit data; in fact its natural mass scale could be anywhere below the Planck mass. It is this aspect of the sterile neutrino physics that we explore in this paper, i.e. how to have a light sterile neutrino with mass in the eV range naturally while also simultaneously explaining the small masses of known active neutrinos. The model we discuss is a variation of a mirror model that was originally proposed to explain a light asymmetric dark matter (ADM)~\cite{ADM-An} and the relation $\Omega_{\rm DM} \simeq 5 \Omega_{\rm baryonic}$; as we will show in this paper, this model naturally leads to a light sterile neutrino with mass in the eV range, as required to fit current anomalies.

A simple way to understand the small sterile neutrino mass is to adopt the same strategy as is generally adopted for the active neutrinos, i.e. to invoke an analog of the $B-L$ symmetry whose breaking at high scale is at the root of seesaw mechanism for small active neutrino masses~\cite{seesaw}. A realization of this proposal can be given by assuming that the SM has a mirror counter part with identical forces and matter, with interactions in the two sectors being related by mirror symmetry~\cite{okun}. The presence of the mirror symmetry prevents the proliferation of coupling parameters that could be feared to accompany the doubling of particle number. These models are also well motivated as providing a candidate for dark matter since the lightest mirror baryon is stable for the same reason as the familiar proton, i.e. the analog of baryon number in the mirror sector, denoted by $B^\prime$. The relevance of this model to the lightness of the sterile neutrino is that just like in the SM, the three neutrinos of the mirror sector are massless due to the mirror analog of $B-L$ symmetry (denoted by $(B'-L')$ henceforth)~\cite{mirrorsterile}. A mirror analog of the seesaw mechanism could provide a natural way for sterile neutrinos to be light, making them perfect candidates for being sterile neutrinos. It, however, turns out that when the requirement of getting a dark matter out of this theory as in~\cite{ADM-An} is imposed, what is required is a variant of the usual seesaw mechanism, where we use a common set of heavy Majorana right-handed neutrinos connecting the visible and the mirror sector, instead of two sets as would be the case if we na\"ively implement the seesaw mechanism in both sectors. Since the right-handed neutrinos are SM singlet fields, their couplings to lepton doublets and Higgs doublets of each sector are allowed by full gauge invariance of the theory. We describe the details of this model below but to summarize, the leading order result of adding a common set of right-handed neutrinos is that the neutrino mass matrix is a $(6+n_R)\times (6+n_R)$ matrix (where $n_R$ is the number of right-handed neutrinos), having the inverse seesaw form~\cite{inverse}. It turns out that when we impose the constraints of ADM in this model a la~\cite{ADM-An}, it predicts too large a mass (in the 100s of MeV) for mirror neutrinos if $n_R = 3$. However, if we choose $n_R=2$, one of the three sterile neutrinos remains massless at tree level and picks up a small eV level mass due to Planck scale effects. This plays the role of the eV sterile neutrino. We then show how one-loop effects generate the masses and mixings of the active neutrinos as well as the mixings between the active and sterile neutrinos. We find an interesting result that to get the active-sterile mixing (ASM), we need a two Higgs doublet extension of the standard as well as mirror model.

The paper is organized as follows: in Sec. II, we present an outline of the ADM model which provides a setting for the sterile neutrino discussion; in Sec. III, we present the neutrino sector of the model and show how the 3+1 model arises. Sec, IV is devoted to a detailed investigation of the neutrino masses and mixings in the active sector as well as the cross mixings between the sterile and active neutrinos. Simple numerical examples are shown in Sec. V before we conclude in Sec. VI.

\section{Basic outline of the mirror ADM model}
In this section, we review the basic features of the mirror ADM model of Ref.~\cite{ADM-An}, which gives an explanation of the similar magnitudes of dark matter content and baryon content of the Universe. The basic postulate of the model is the existence of a mirror sector analogical to our universe with identical forces and matter content, i.e. the gauge groups of the mirror sector is $SU(3)'_c\times SU(2)'_L\times U(1)'_Y$ and the matter and Higgs assignments in the mirror sector are identical to those in the familiar SM sector. For simplicity, the mirror partners of the SM gauge bosons and fermions will be denoted by prime over the corresponding SM fields. Thus each particle that we know has a mirror partner. The two sectors are always connected by gravity, which couples to all energy and matter universally. The model has a mirror symmetry ( $Z_2$ symmetry) which connects all the interactions and couplings of the two sectors. Thus there are no new parameters in the model prior to symmetry breaking, even though the particle spectrum is doubled. Symmetry breaking is chosen such that the mirror electroweak vacuum expectation value (VEV) $v'_{wk} \gg v_{wk}$. How this can be done starting with an exact mirror symmetric Hamiltonian has been explained in Ref.~\cite{dolgov}. In this paper, we follow the basic framework outlined in Ref.~\cite{ADM-An}; however, for one of the crucial parameter of the model, the ratio $v'_{wk}/v_{wk}$, we choose a higher value, i.e. $v'_{wk}/ v_{wk}\sim 10^4$ rather than $10^3$ chosen in Ref.~\cite{ADM-An} to ensure that the heavier mirror neutrinos decay before the BBN epoch (see the discussion later). Due to this asymmetry, the mirror quarks become $10^4$ times heavier than the familiar SM quarks. However, depending on the nature of symmetry breaking in the mirror sector compared to the familiar sector (e.g. if there are two Higgs doublets with different mirror asymmetric VEVs and hence a different value of tan$\beta$ in the mirror sector), the mirror quarks could be slightly lighter than $10^4\, m_q$. If we then assume that at some super heavy scale the gauge couplings unify (or similar in magnitude), then, due to earlier decoupling of the mirror quarks from the mirror strong coupling evolution (compared to the familiar QCD coupling evolution), we will have~\cite{ADM-An} $\Lambda'_{\rm QCD} \gg \Lambda_{\rm QCD}$, e.g. with $\Lambda'_{QCD}\simeq 2$ GeV. It was argued in Ref.~\cite{ADM-An} that this could lead easily to the lightest mirror baryon (mirror neutron) with $m_{B'}\sim 5 m_B$\footnote{In the present work with $v'_{wk}/v_{wk} \sim 10^4$ which is different from the model in~\cite{ADM-An}, we need a deep gluon potential to cancel out the large mirror quark masses, or other extra ingredients, to obtain a $\sim$5 GeV dark matter. More detailed discussion on this issue is given in Appendix A.}, thereby explaining the dark matter to baryonic mass density relation: $\Omega_{\rm DM} \simeq 5 \Omega_{\rm baryonic}$.

The next issue in such models is how to satisfy the Big Bang Nucleosynthesis (BBN) constraints since there is a duplication of neutrino and photon number in the model. In Ref.~\cite{ADM-An}, a symmetry breaking pattern is chosen such that all the mirror neutrinos are much heavier and decay before the BBN temperature. Similarly, the mirror photon is chosen to have a mass of about 100 MeV so that via its mixing with familiar photon, it also decays before the BBN epoch. Here we will take a slightly different path as we explain below. Roughly speaking, we will assume that BBN allows an effective neutrino number to be four so that we can take one light sterile neutrino with about 10\% mixing with the active neutrinos, as suggested by experiments. The two other mirror particles that could create problems for BBN are: (i) the mirror photon and (ii) the mirror electron, whose mass is about 5 GeV in our model. The mirror photon will be assumed massive as in Ref.~\cite{ADM-An} as just noted and will be allowed to decay via $\gamma-\gamma'$ mixing; as was shown in~\cite{ADM-An}, this will decay safely before the BBN epoch to $e^+e^-$. As far as the mirror electron is concerned, $e^{\prime +}e^{\prime -}$ can annihilate to $\gamma'\gamma'$ states and depleted enough to make a negligible contribution to the energy density of the universe at the BBN epoch. As shown in Appendix B, for low GeV range of electron mass, indeed this appears to be the case. Alternatively, we may adopt the same strategy as in~\cite{ADM-An} and assume that we break the mirror electric charge by a non-zero VEV to the charged member of the second mirror Higgs doublet which couples to the heavier mirror leptons. This not only gives mass to the mirror photon just discussed but also lets $e'\to \nu'+\gamma'$ with a lifetime much less than a second so as not to affect BBN.  All the details of this discussion are given in ~\cite{ADM-An}. In any case, these aspects of the model are not pertinent to the discussion of the paper but are mentioned only to point out that the overall model  is quite compatible with cosmology.

\section{Neutrino sector and tree level neutrino spectrum}

The neutrino sector of the model is the key focus of this paper.  We therefore identify explicitly some of the key points and quantities relevant to this discussion. Unlike Ref.~\cite{ADM-An}, we introduce only two heavy right-handed neutrinos $N_i$ to connect the ordinary and mirror sectors, or specifically the three flavors of active neutrinos $\nu$ as well as mirror neutrinos $\nu^{\prime}$ are connected in a mass matrix via the two right-handed ``bridge'' neutrinos. To proceed with further analysis, we first note that we can parameterize the right-handed mass matrix as a diagonal one, by a suitable choice of basis, i.e.
\begin{eqnarray}
M_R = \left(\begin{matrix}
  M_1 & 0 \\ 0 & M_2
\end{matrix}\right) \,.
\end{eqnarray}
The Yukawa coupling involving the two right-handed neutrinos connecting the two sectors can be written as:
\begin{eqnarray}
{\cal L}^\nu_Y= (\vec{h}_1 \cdot \vec{\bar{L}}\phi + \vec{h}_1 \cdot \vec{\bar{L}}^\prime\phi') {N}_1
+ (\vec{h}_2 \cdot \vec{\bar{L}}\phi + \vec{h}_2 \cdot \vec{\bar{L}}^\prime\phi') {N}_2 + {\rm h.c.} \,.
\end{eqnarray}
Generally $\vec{h}_1= (h_{11}, h_{12}, h_{13})$ and $\vec{h}_2=(h_{21}, h_{22}, h_{23})$, and then the resultant Dirac mass matrix and its mirror counterpart read, respectively,
\begin{equation}
\begin{array}{ll}
M_D = \left(\begin{matrix}
  a & b \\ c & d \\ e & f
\end{matrix}\right) \,, &
M'_D = \left(\begin{matrix}
  A & B \\ C & D \\ E & F
\end{matrix}\right) \,.
\end{array}
\end{equation}

As there are only two right-handed neutrinos in the model, by a appropriate unitary transformation, we can always decouple the two massless states (at tree level)
\begin{eqnarray}
&& \widetilde{\nu}_\tau \propto 
\left( \frac{-de+cf}{ad-bc}\nu_e +\frac{be-af}{ad-bc}\nu_\mu +\nu_\tau \right) \,, \\
&& \widetilde{\nu}_\tau^\prime \propto 
\left( \frac{-DE+CF}{AD-BC}\nu'_e +\frac{BE-AF}{AD-BC}\nu'_\mu +\nu'_\tau \right) \,,
\end{eqnarray}
from the other massive ones, leaving the simplified $2\times2$ (mirror) Dirac matrix connecting the two left-handed neutrinos $\widetilde{\nu}_{e,\,\mu}$ ($\widetilde{\nu}^\prime_{e,\,\mu}$) and the heavy right-handed ones, written in the form of, respectively,
\begin{equation}
\begin{array}{ll}
\widetilde{M}_D = \left(\begin{matrix}
  \tilde{a} & \tilde{b} \\ 0 & \tilde{d}
\end{matrix}\right) \,, &
\widetilde{M}_D^\prime = \left(\begin{matrix}
  \tilde{A} & \tilde{B} \\ 0 & \tilde{D}
\end{matrix}\right) \,.
\end{array}
\end{equation}
Consequently, the $6\times 6$ neutrino mass matrix is, in the basis of $(\tilde{\nu}_{e,\,\mu},\, N^C,\, \tilde{\nu}^\prime_{e,\,\mu}$),
\begin{eqnarray}
\label{matrix}
\widetilde{M}_\nu = \left(\begin{matrix}
0 & \widetilde{M}_D & 0 \\
\widetilde{M}_D^T & M_R & \widetilde{M}_D^{\prime T} \\
0 & \widetilde{M}_D^\prime & 0
\end{matrix}\right) \,.
\end{eqnarray}
In this case, there exist two more massless states at tree level, which are in the form of $\nu_i \propto (v'_{wk}\tilde {\nu}_i-v_{wk}\tilde{\nu}'_i)$ ($i=1,2$) where $\tilde{\nu}_i$ and $\tilde{\nu}_i^\prime$  have identical flavor structure prior to seesaw diagonalization.  It follows from this expression that $\nu_i$ are predominantly active neutrino states. Summarizing then, we have a total of three active neutrino states and one sterile state which are massless at tree level. This is the beginning of the construction of the $3+1$ model. Next we discuss how these various states pick up masses.

Two of the three active massless states obtain their masses radiatively~\cite{MRISM1}, details of which are described in the subsequent section. The magnitude of those masses are, to the zeroth order in ${m^2_{H,\,Z}}/{M^2_R}$,
\begin{eqnarray}
\label{mnu}
m_\nu \sim \frac{g^2}{16\pi^2} \Big( M_D M_R^{-1} M_D^T \Big)  \ln\left(\frac{M^2_R}{v^2_{wk}}\right) \,,
\end{eqnarray}
via the scalar and $Z$ boson loops. Of the three sterile neutrinos, the two massive ones pick up mass via tree level seesaw as follows:
\begin{eqnarray}
\label{mnu'}
m_{\nu'} \sim  M'_D M_R^{-1} M_D^{\prime T} \,.
\end{eqnarray}
Since the model of Ref.~\cite{ADM-An} requires for reasons of leptogenesis that $M_R\sim 10^8$ GeV, we keep this value here and using the assumption stated earlier that $v'_{wk}\sim 10^4\, v_{wk}$, we  find that the heavier sterile neutrino  masses are of order $m_{\nu'}\sim 10^4 h^2_\nu$ GeV. For $h_\nu \sim 0.003-0.01$, we get $m_{\nu'}\sim 100-1000$ MeV. With reasonable mixing with active neutrinos, these sterile states have short enough lifetime to decay before the BBN epoch. We now turn to a discussion of the mass of the lightest sterile neutrino. The main source of its mass seems to be the gravitationally induced $d=5$ gauge invariant interaction~\cite{Weinberg-d=5}
\begin{eqnarray}
\label{sn-operator}
\frac{1}{M_{\rm pl}} \Big[ ( L_\tau \phi)^2 + ( L'_\tau \phi')^2 \Big] \,,
\end{eqnarray}
where $M_{\rm pl}$ is the Planck scale. The ASMs of order 10\% are radiatively generated via the mixing between the scalars in two sectors as we will see in a later section. With all the ingredients above, we arrives at the naturally built model for the 3+1 scenario.

\section{Implementation of the 3+1 model}
In this section we demonstrate explicitly all the ingredients aforementioned, including radiative generation of the active neutrino masses and mixings, gravitational generation of the eV scale mass, as well as the mixings between the active and sterile neutrinos induced by mixing of the scalars in the two sectors of the ADM model.

\subsection{Radiative mass generation of the active neutrinos via inverse seesaw}
\begin{figure}[tp]
  \centering
  \includegraphics[width=7cm]{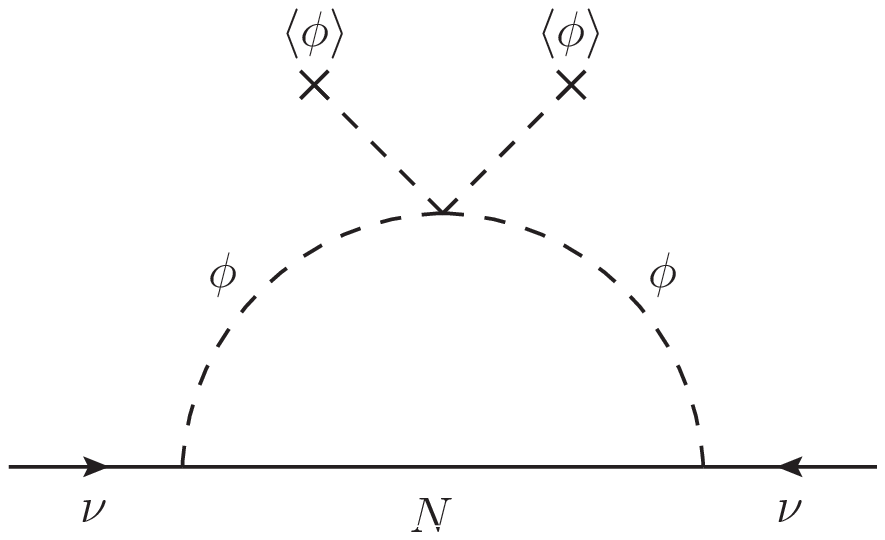}  \hspace{-.5cm}
  \includegraphics[width=7cm]{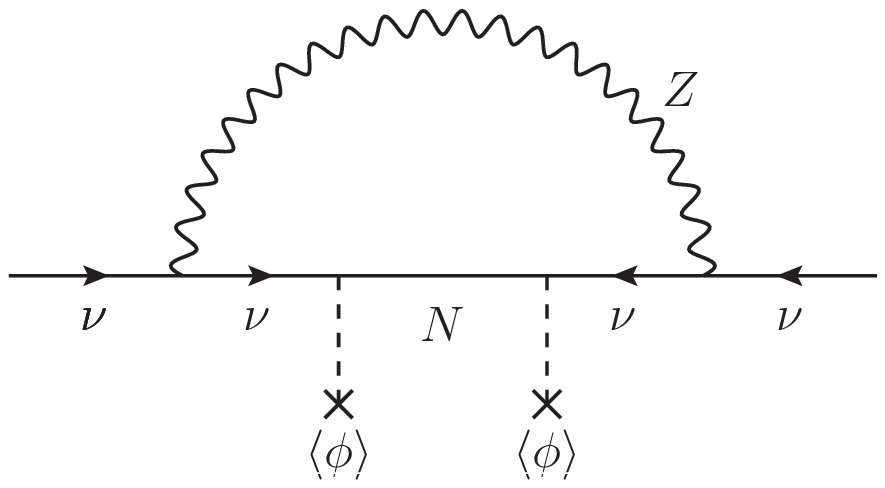}
  \caption{Radiative generation of the active neutrino masses via inverse seesaw by the SM Higgs and $Z$ boson loops~\cite{MRISM1}.}
  \label{fig:nu}
\end{figure}

In the ordinary neutrino sector, the two massless states in Eq.~(\ref{matrix}), predominately active neutrinos, can obtain their masses radiatively via inverse seesaw~\cite{MRISM1}, i.e. by the SM Higgs and $Z$ boson loops shown in Fig.~\ref{fig:nu}. Using the Feynman rules for flavor number violating interactions~\cite{FNV}, we can obtain the effective neutrino mass at 1-loop level,
\begin{eqnarray}
(M_{\nu})_{\alpha\beta} &=& \sum_{\gamma} \Big[ f_{\alpha\beta\gamma}(m_H) +3f_{\alpha\beta\gamma}(m_Z) \Big] \,,
\end{eqnarray}
where we have defined the function
\begin{eqnarray}
f_{\alpha\beta\gamma}(m) \equiv
\frac{\alpha_W}{16\pi m_W^2} (M_D)_{\alpha\gamma} (M_R)_{\gamma\gamma}
 \left[ \frac{m^2}{(M_R^2)_{\gamma\gamma}-m^2} \ln\left(\frac{(M_R^2)_{\gamma\gamma}}{m^2}\right) \right] (M_D)_{\beta\gamma} \,,
\end{eqnarray}
with $\alpha_W=g^2/4\pi$ is the weak coupling constant, $m_W$, $m_H$ and $m_Z$ respectively the masses of the SM $W$, $H$ and $Z$ bosons, and $\alpha$, $\beta$ and $\gamma$ the indices for the matrices $M_D$ and $M_R$. In the limit of $M_1 = M_2$\footnote{In our ADM model, we choose the option of resonant leptogenesis which requires $M_1 \simeq M_2$ as in Ref.~\cite{ADM-An}.}, the effective active neutrino mass matrix can be factorized as
\begin{eqnarray}
\label{eqn:Mnu}
M_{\nu} = \frac{\alpha_W}{16\pi}
\left( M_D M_R^{-1} M_D^T \right)
 \left[ \frac{ m^2_H/m^2_W}{1-m^2_H/M^2} \ln\left(\frac{M^2}{m^2_H}\right)
+       \frac{3m^2_Z/m^2_W}{1-m^2_Z/M^2} \ln\left(\frac{M^2}{m^2_Z}\right) \right]\,,
\end{eqnarray}
where $M=M_{1,\,2}$, and the seesaw factor
\begin{eqnarray}
\label{eqn:Mnu2}
M_D M_R^{-1} M_D^T = \frac1M
\left(\begin{array}{ccc}
 a^2+b^2 & a c+b d & a e+b f \\
 a c+b d & c^2+d^2 & c e+d f \\
 a e+b f & c e+d f & e^2+f^2 \\
\end{array}\right) \,.
\end{eqnarray}
As we expect, the masses of the three active neutrinos are suppressed by both the seesaw mechanism and loop factor. Moreover, due to the fact that $M_D M_R^{-1} M_D^T$ is of rank two, one of the active neutrinos $\tilde{\nu}_\tau$ remains massless even at one-loop order. This could be either $m_1$ or $m_3$ depending on whether we have normal or inverted hierarchy (see our numerical analysis below). However, we stress that generally the reactor mixing angle $\theta_{13}$ does not vanish.

When the scalar sectors are extended with two or more Higgs doublets, the massless active state $\tilde{\nu}_\tau$ can generally become massive. However, in the limit that the contributions from heavy scalars and pseudoscalars cancel out each other, we are left with the ``leading order'' SM contributions in Eq.~(\ref{eqn:Mnu}).

\subsection{ Sterile neutrino mass from gravitational interactions}

In an analogous manner to the one-loop generation of the active neutrino masses, the 1-loop diagrams with the mirror particles and the right-handed neutrinos would generate corrections to the mirror neutrino masses, but they do not affect the state $\widetilde{\nu}_\tau^\prime$ which decouples completely from other states, as long as the heavier (pseudo)scalars decouple from the lighter one as it is in the visible sector. We postulate that the eV sterile neutrino mass required for the SBL experiments is generated from gravitational interactions, e.g. via the dimension-5 operator in Eq.~(\ref{sn-operator}). For the active neutrinos, the masses from the the dimension-5 operators are extremely small, mostly of order $10^{-6}$ eV, and can be safely neglected. However, for the mirror neutrinos, after the mirror scalar get a non-vanishing VEV, we can obtain the effective operator
\begin{eqnarray}
\frac{v^{\prime 2}}{M_{\rm pl}} \overline{\tilde{\nu}_\tau^\prime} \tilde{\nu}_\tau^{\prime C} \,.
\end{eqnarray}
With a coefficient of about $5\times10^{-3}$, we arrive at the desired eV scale mass.\footnote{In Eq.~(\ref{sn-operator}) we did not consider the cross terms like $(L_\tau \phi)(L'_\tau \phi')/ M_{\rm pl}$. After the scalars get non-vanishing VEVs, their contribution to the ASMs can be safely ignored: Even given an order one coefficient, the generated mass is about of order 0.01 eV and not large enough (up to about 0.1 eV) to explain the SBL experiments.}

\subsection{Radiative generation of the active-sterile mixings}

\begin{figure}[tp]
  \centering
  \includegraphics[width=7cm]{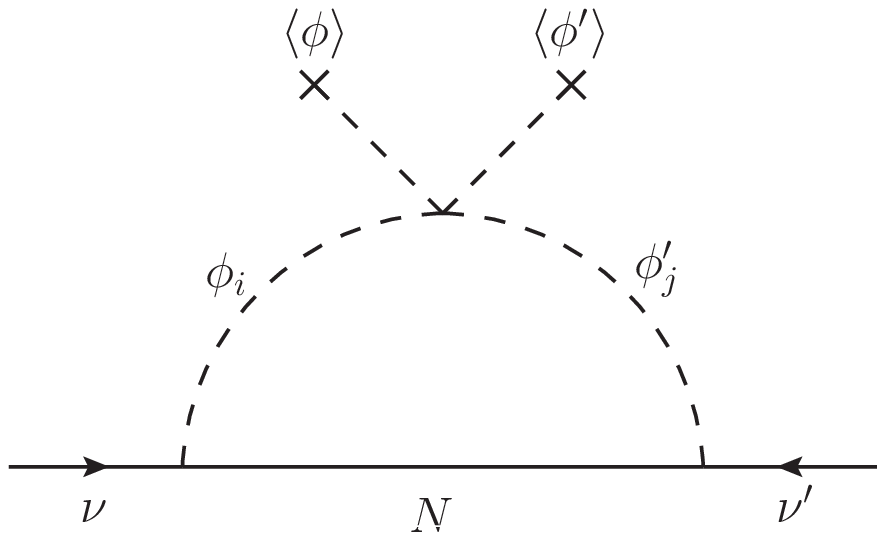}  \hspace{-.5cm}
  \includegraphics[width=7cm]{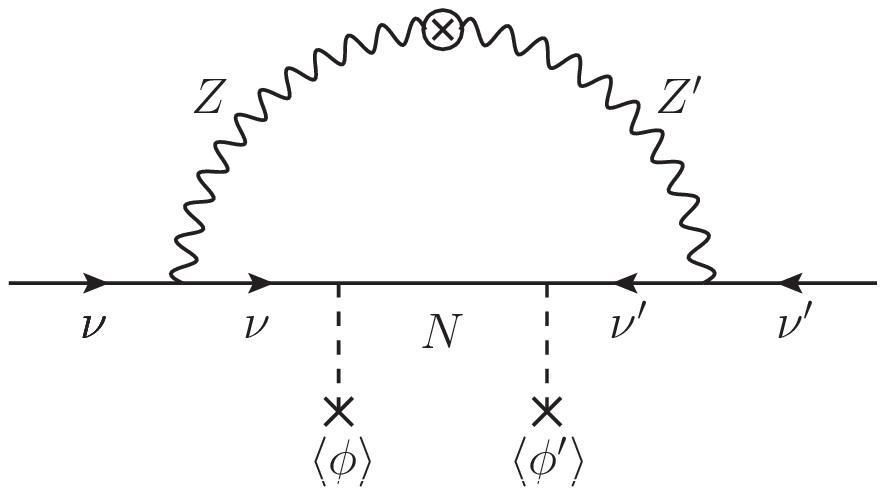}
  \caption{Mixing between the active and mirror neutrinos via the loops induced by the mixing between the scalars $\phi$ and $\phi^\prime$ and the kinetic $Z - Z^\prime$ mixing.}
  \label{fig:mixing}
\end{figure}

Besides the right-handed neutrinos and the gravitational interactions, the ordinary and mirror sectors in the ADM model can also be connected by couplings of the scalars and mixing of the gauge bosons. With regard to the ASMs, the relevant connections are the kinetic mixing of the ordinary and mirror $Z$ bosons and the four-scalar interactions in the potential
\begin{eqnarray}
\label{mixing-Lagrangian}
\mathcal{V} \supset \sum_{ijkl} x_{ijkl} (\phi^\dagger_i\phi_j) (\phi^{\prime \dagger}_k \phi^\prime_l) \,.
\end{eqnarray}
Both these two channels contribute via one-loop diagrams, as shown in Fig.~\ref{fig:mixing}.
For the $Z - Z^\prime$ mixing loop, it is reasonable to assume that the $Z^\prime$ boson couples universally to the mirror neutrinos, as the weak interaction dedicates in the real world. Then, in this case, the $Z - Z^\prime$ mixing is proportional to $M_D M_R^{-1} M_D^{\prime T}$ and does not actually contribute to the ASMs; thus we need only to consider the scalar loops to produce the $\sim$10\% ASMs.

For simplicity, we first consider the case with only one Higgs doublet in both the viable and mirror sectors, and we will see explicitly why we need at least two Higgs doublets to generate the non-vanishing ASMs, and why the $Z - Z'$ mixing does not induce the ASMs as just mentioned. When the $\phi$ and $\phi'$ get non-vanishing VEVs, we arrive at the mixing between the normal and mirror Higgs scalars,
\begin{eqnarray}
\mathcal{V} \supset x vv' HH^{\prime} \,,
\end{eqnarray}
then the resultant mixing mass matrix reads
\begin{eqnarray}
(M_{\nu\nu^\prime})_{\alpha\beta} = x \sum_\gamma g_{\alpha\beta\gamma}(m_{H},m'_{H}) \,,
\end{eqnarray}
where we have defined the function
\begin{eqnarray}
g_{\alpha\beta\gamma}(m_H,m'_H) &\equiv&
\frac{1}{16\pi^2 (m^{\prime 2}_H-m^2_H)} (M_D)_{\alpha\gamma} (M_R)_{\gamma\gamma} \nonumber \\
&&  \left[ \frac{m^{\prime 2}_H}{(M_R^2)_{\gamma\gamma}-m^{\prime 2}_H} \ln\left(\frac{(M_R^2)_{\gamma\gamma}}{m^{\prime 2}_H}\right)
          -\frac{m^2_H}{(M_R^2)_{\gamma\gamma}-m^{\prime 2}_H} \ln\left(\frac{(M_R^2)_{\gamma\gamma}}{m^2_H}\right)  \right] (M'_D)_{\beta\gamma} \,. \nonumber \\
\end{eqnarray}
As the scalars in the mirror sector is much heavier than those in the ordinary sector, we can safely make the approximation
\begin{eqnarray}
g_{\alpha\beta\gamma}(m'_H) &\simeq&
\frac{1}{16\pi^2 m^{\prime 2}_H} (M_D)_{\alpha\gamma} (M_R)_{\gamma\gamma}
\left[ \frac{m^{\prime 2}_H}{(M_R^2)_{\gamma\gamma}-m^{\prime 2}_H} \ln\left(\frac{(M_R^2)_{\gamma\gamma}}{m^{\prime 2}_H}\right) \right] (M'_D)_{\beta\gamma} \,,
\end{eqnarray}
which is independent of the ordinary scalar masses. Analogical to Eq.~(\ref{eqn:Mnu}), with $M_1=M_2$, the mixing mass matrix can be factorized into the form
\begin{eqnarray}
\label{eqn:mixing}
M_{\nu\nu'} = \frac{x}{16\pi^2}
\left( M_D M_R^{-1} M_D^{\prime T} \right)
\left[ \frac{1}{1-m^{\prime2}_H/M^2} \ln\left(\frac{M^2}{m^{\prime2}_H}\right) \right]\,,
\end{eqnarray}
with the seesaw factor
\begin{eqnarray}
\label{eqn:mixing2}
M_D M_R^{-1} M_D^{\prime T} = \frac1M
\left(\begin{array}{ccc}
 aA+bB & aC+bD & aE+bF \\
 cA+dB & cC+dD & cE+dF \\
 eA+fB & eC+fD & eE+fF \\
\end{array}\right) \,.
\end{eqnarray}
This form of $M_{\nu\nu'}$ has the property that
\begin{eqnarray}
M_{\nu\nu'}|\widetilde{\nu}_\tau^\prime\rangle=0 \,,
\end{eqnarray}
which implies that the light sterile neutrino does not mix with the active ones! Similarly, it is clear that the contribution from $Z - Z'$ loop is also proportional to $M_D M_R^{-1} M_D^{\prime T}$ and therefore cannot generate the ASMs. Consequently, in the absence of the Planck scale induced effects, the two massless states ($\widetilde{\nu}_\tau$ and $\widetilde{\nu}_\tau^\prime$) are protected from radiative electroweak corrections in models with the minimal scalar sector. However, when the scalar sector is extended to include more doublets, due to the two sets of independent Yukawa couplings, $M_{\nu,\nu'}$ is not proportional to $M_D M_R^{-1} M_D^{\prime}$ anymore and ASMs can be generated.\footnote{The extra heavy neutral scalars and pseudoscalars may lead to flavor changing neutral currents beyond the SM, e.g. the $\mu \rightarrow e \gamma$ rare decay process. Following~\cite{FCNC}, if we assume the flavor changing Yukawa couplings are proportional to $\sqrt{m_im_j}$ with $m_{i,\,j}$ the masses of charged leptons involved and the the heavy scalar contribution dominates, the MEG experiment with a sensitivity of $5.7\times10^{-13}$~\cite{MEG} would lead to a lower limit of about 350 GeV on the heavy scalar mass.} To estimate the magnitude of this mixing, we can parameterize the effective mixing matrix as follows:
\begin{eqnarray}
M_{\nu\nu'}^{\rm eff} = \frac{x}{16\pi^2}
\left[ \frac{1}{1-m^{\prime2}_H/M^2} \ln\left(\frac{M^2}{m^{\prime2}_H}\right) \right]
\cdot \frac{1}{M} \left(\begin{matrix} c_1 kK \\ c_2 kK \\ c_3 kK \end{matrix}\right) \,,
\end{eqnarray}
where $k$ and $K$ are, respectively, the mass scales of the matrix elements of $M_D$ and $M'_D$, and $c_i$ coefficients of order 0.1 indicating the cancellation between the scalar and pseudoscalar loops. The mixing matrix elements are typically of order $\frac{h^2_\nu}{16\pi^2}xc_i\frac{v_{wk}v'_{wk}}{M_R}\sim 10^{-7}x$ GeV for $h_\nu\sim 10^{-2}$, $M_R\sim 10^8$ GeV; for $x\sim 10^{-3}$ They produce $\nu-\nu'$ mixings of order 10\%.

\section{Numerical example}
\label{sec:numrical}
In this section, we give several numerical examples to illustrate the point that our model can reproduce the observed mixings for the three neutrino sub-sector while accommodating an eV sterile neutrino with mixing of order 10\% for reasonable choice of parameters. Obviously, these are not predictions; instead they are meant to provide a guide as to expectations for observables such as neutrinoless double beta decay, that we discuss in the last subsection. We fit the parameters of the model so as to reproduce the observed oscillation parameters for the active neutrinos and present fits for the 3+1 model~\cite{nu-fit,sn-fit-Giunti1,sn-fit-theta34} (see Tables I and II). Due to the large $\tau^-$ mass, the SBL experiments are not sensitive to the tauon flavor, thus the constraint on $\theta_{34}$ is comparatively rather loose, $0 < \theta_{34} < 30^\circ$; in our fit, as an illustrative example, we set explicitly the best fit value $\theta_{34}=5^\circ$.
\begin{table}[tp]
  \centering
  \caption{Experimental data on neutrino oscillation parameters~\cite{nu-fit}.}
  \label{table:nu-fit}
  \begin{tabular}{cccc}
  \hline\hline
    Parameter & Hierarchy & Best fit  & 3$\sigma$ range    \\ \hline
    $\Delta m^2_{21}\; [10^{-5}\;\text{eV}^2]$  & NH/IH & 7.62          & $7.12 - 8.20$ \\ \hline
    $|\Delta m^2_{31}|\; [10^{-3}\;\text{eV}^2]$       &NH     & 2.55          & $2.31 - 2.74$ \\
                                              &IH     & 2.43          & $2.21 - 2.64$ \\ \hline
    $\theta_{12}$                             & NH/IH & $34.4^\circ$  & $31.3^\circ - 37.5^\circ$ \\ \hline
    $\theta_{23}$                             & NH    & $51.5^\circ$  & $36.9^\circ - 55.6^\circ$ \\
                                              & IH    & $50.8^\circ$  & $37.5^\circ - 54.9^\circ$ \\ \hline
    $\theta_{13}$                             & NH    & $9.0^\circ$   & $7.5^\circ - 10.5^\circ$ \\
                                              & IH    & $9.1^\circ$   & $7.5^\circ - 10.5^\circ$ \\ \hline\hline
  \end{tabular}
\end{table}
\begin{table}[tp]
  \centering
  \caption{Experimental data on sterile neutrino parameters~\cite{sn-fit-Giunti1,sn-fit-theta34}. See text for details. }
  \label{table:sn-fit}
  \begin{tabular}{cccc}
  \hline\hline
    Parameter & Hierarchy & Best fit & 3$\sigma$ range    \\  \hline
    $\Delta m^2_{41} \; [\text{eV}^2]$  & NH/IH  & 1.62  & $0.72 - 2.53$ \\ \hline
    $\theta_{14}$                       & NH/IH  & $10.1^\circ$  & $6.1^\circ - 15.3^\circ$ \\ \hline
    $\theta_{24}$                       & NH/IH  & $ 5.9^\circ$  & $2.8^\circ - 12.5^\circ$ \\ \hline
    $\theta_{34}$                       & NH/IH  & $ 5.0^\circ$  & $0^\circ - 30^\circ$ \\ \hline\hline
  \end{tabular}
\end{table}

In the model I below, the active neutrinos have a normal mass hierarchy whereas in model II, the mass hierarchy is inverted. We also present a model (model III) where the active neutrino masses are quasi-degenerate.

\subsection{Model I with normal hierarchy}
If we choose the following parameter values (in unit of GeV)
\begin{equation}
\begin{array}{lll}
a=  0.0157 + 0.216 i \,, &
b=  0.305  - 0.0112i \,, &
c= -0.152  + 0.171 i \,, \\
d=  0.241  + 0.108 i \,, &
e= -0.163  + 0.125 i \,, &
f=  0.177  + 0.115 i \,,
\end{array}
\end{equation}
and (with $x=1.5 \times 10^{-3}$, $m_{H'}=126\times10^4$ GeV, $k=0.33$ GeV, and $K=0.33\times 10^4$ GeV)
\begin{equation}
\begin{array}{lll}
c_1 = 0.306 \,, &
c_2 = 0.167 \,, &
c_3 = 0.144 \,,
\end{array}
\end{equation}
we arrive at the $4\times4$ effective neutrino mass matrix\footnote{Due to active-sterile mixing, the $3\times3$ submatrix of the matrix (\ref{N1}) can generally no longer be cast into the form of Eq.~(\ref{eqn:Mnu2}). That is to say, we need to include the ``subleading'' corrections of the heavy (pseudo)scalars to the active neutrino mass matrix which would spoil the rank=2 matrix $M_D M_R^{-1} M_D^{T}$. In this example the correction is very small, i.e. a correction of (negative) 0.0024 ${\rm GeV}^2$ to the (non-)diagonal elements of the matrix $M_D M_D^T$.}
\begin{eqnarray}
\label{N1}
\left(\begin{array}{cccc}
 0.0541 & 0.0366 & 0.0256 & 0.277 \\
 0.0366 & 0.0475 & 0.0345 & 0.152 \\
 0.0256 & 0.0345 & 0.0346 & 0.131 \\
 0.277 & 0.152 & 0.131 & 1.54 \\
\end{array}\right) \, {\rm eV} \,.
\end{eqnarray}
with normal hierarchy $m_1 \simeq0 \ll m_2 \ll m_3$. Although $m_1$ may mix with the sterile flavor as well as other active states, a massless state $m_1\simeq0$ is permitted in our model.

\subsection{Model II with inverted hierarchy}
Similarly, we can also have a fit for the IH case with a massless $m_3$, if we choose these parameters as follows (in unit of GeV),
\begin{equation}
\begin{array}{lll}
a=  0.506  + 0.303  i \,, &
b=  0.339  - 0.452  i \,, &
c=  0.318  + 0.0069 i \,, \\
d=  0.0077 - 0.284  i \,, &
e=  0.0568 + 0.314  i \,, &
f=  0.352  - 0.0507 i \,,
\end{array}
\end{equation}
and
\begin{equation}
\begin{array}{lll}
c_1 = 0.299 \,, &
c_2 = 0.176 \,, &
c_3 = 0.150 \,,
\end{array}
\end{equation}
which result in the effective neutrino mass matrix
\begin{eqnarray}
\label{N2}
\left(\begin{array}{cccc}
 0.0970 & 0.0220 & 0.0186 & 0.271 \\
 0.0220 & 0.0368 & -0.0097 & 0.159 \\
 0.0186 & -0.0097 & 0.0426 & 0.136 \\
 0.271 & 0.159 & 0.136 & 1.54 \\
\end{array}\right) \, {\rm eV} \,.
\end{eqnarray}

\subsection{Model III with quasi-degenerate active neutrinos}
In our model, it is also possible that the lightest neutrino $m_1$ or $m_3$ obtain a non-vanishing value from mixing with the sterile flavor, e.g. $m_1 \simeq m_4 \sin^2\theta_{14}$. Then in this case the three active neutrinos are mostly likely to be quasi-degenerate. Here follows one example: given the parameter set with the values (in unit of GeV)
\begin{equation}
\begin{array}{lll}
a= 0.0195 \,, &
b= 0.258  \,, &
c= 0.114 \,, \\
d= 0.187 \,, &
e= 0.106 \,, &
f= 0.170 \,,
\end{array}
\end{equation}
and
\begin{equation}
\begin{array}{lll}
c_1 = 0.297 \,, &
c_2 = 0.166 \,, &
c_3 = 0.142 \,,
\end{array}
\end{equation}
we get the $4\times4$ effective neutrino mass matrix
\begin{eqnarray}
\label{N3}
\left(\begin{array}{cccc}
 0.0996 & 0.0303 & 0.0252 & 0.269 \\
 0.0303 & 0.0784 & 0.0229 & 0.151 \\
 0.0252 & 0.0229 & 0.0697 & 0.129 \\
 0.269 & 0.151 & 0.129 & 1.54 \\
\end{array}\right) \; {\rm eV} \,.
\end{eqnarray}

\subsection{Expectations for neutrinoless double beta decay ($\beta\beta_{0\nu}$) }
\begin{figure}
  \centering
  \includegraphics[width=7.7cm]{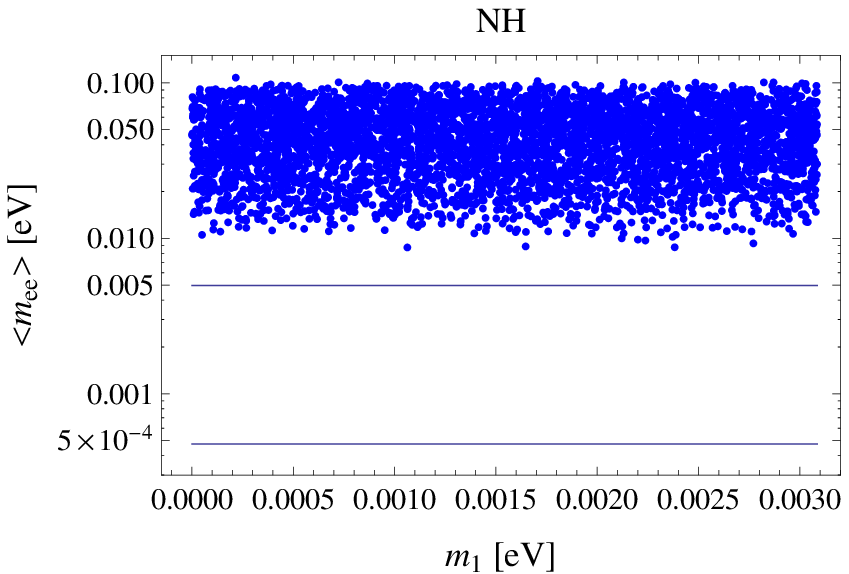} \hspace{-.3cm}
  \includegraphics[width=7.7cm]{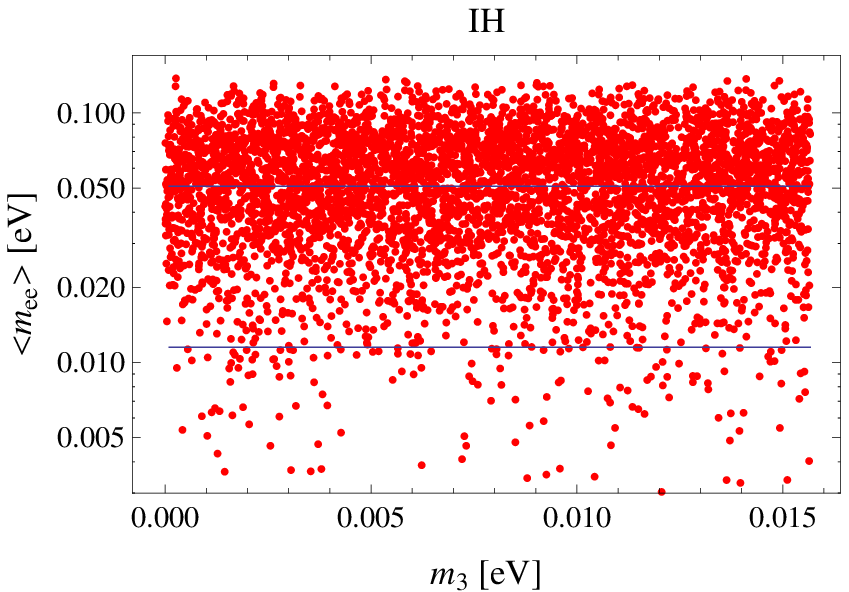}
  \caption{Predictions of our model on the effective neutrino mass $\langle m_{ee} \rangle$ in neutrinoless double beta decay. Left panel: the NH case with $m_1 \ll m_2 \ll m_3$. Right panel: the IH case with $m_3 \ll m_1 \sim m_2$. The horizontal lines indicate the upper and lower limits of $\langle m_{ee} \rangle$ in the three neutrino framework with a massless state (massless $m_1$ for NH and massless $m_3$ for IH).}
  \label{fig:nuless}
\end{figure}

From the mass matrices in Eq.~(\ref{N1}), (\ref{N2}) and (\ref{N3}), we see that the effective mass in neutrinoless double beta decay $\langle m_{ee} \rangle \sim (0.05-0.1)$ eV, which implies a testable signal of our model in current relevant experiments~\cite{nuless}. In Fig.~\ref{fig:nuless} we plot the predicted $\langle m_{ee} \rangle$ in both the NH and IH cases, where all the relevant active and sterile neutrino data are allowed to vary in their $3\sigma$ ranges given in Tables~\ref{table:nu-fit} and \ref{table:sn-fit}. Specially, the $3\sigma$ region for the variables $\Delta m^2_{41}-\theta_{14}$ is from Fig.~1 of Ref.~\cite{sn-fit-Giunti1}.
It is interesting that for the NH case, our model gives a large values of $\langle m_{ee} \rangle$ ($\sim (0.01-0.1)$ eV) in contrast with the pure three active neutrino case. This implies that if neutrinoless double beta decay is discovered in the current round of searches, this does not necessarily imply an inverted neutrino spectrum but could indicate the presence of an eV sterile neutrino. By the same token, if there is no evidence for $\beta\beta_{0\nu}$ decay with effective mass $m_{ee}\geq .01$ eV, a large part of the parameter space of our model will be eliminated. Thus searches for $\beta\beta_{0\nu}$ decay can provide an important test of this model.

\section{Conclusion}

There are some indications of a light sterile neutrino from SBL and possibly cosmological data. Further confirmation of the observations is necessary but at this stage, their theoretical implications and their possible connections to other low energy observations in particle physics have already triggered a lot of activity. This work adds a different speculation in that regard, connecting the dark matter with the sterile neutrinos. We have shown that a recently proposed asymmetric dark matter model leads in a natural manner  to a 3+1 neutrino spectrum which may go part of the way (or all the way ?) to accomodating the current SBL experimental data. Since both the sterile neutrino and the dark matter are beyond the minimal seesaw framework, such connections is quite intriguing and may provide a clue to directions for BSM physics. 
The key ingredients of our model are as follows:
\begin{enumerate}
  \item
  The standard model has an identical counter part to it containing the same matter and force content and these two sectors are always connected by gravity: the so called mirror model. This model has a natural candidate for the dark matter, the lightest baryon of the mirror sector, and three extra neutrinos that can qualify as sterile neutrinos. There are heavy right-handed neutrinos connecting the two sectors that relate the baryon density of both sectors and predict a light dark matter with mass in few GeV range.
  \item
  Due to the fact that we introduce two right-handed neutrinos connecting the two sectors, the three (predominately) active neutrinos and one sterile neutrino are massless at tree level. The masses and mixings of the active neutrinos are generated at 1-loop level via the minimal radiative inverse seesaw mechanism explaining the ultra-lightness of the familiar neutrinos. This is the key feature of our model. Specially, the reactor mixing $\theta_{13}$ is produced in this process. 
  \item
  The eV scale mass for the massless sterile neutrino is generated from the $d=5$ operators (cf. Eq.~(\ref{sn-operator})), which is induced from gravitational interactions and thus highly suppressed by the Planck scale. To avoid the BBN constraint, the other two mirror neutrinos are set at the $\sim$ 100 MeV scale by appropriate choice of the mirror weak scale.
  \item
  We have also analysed in detail the radiative generation of active-sterile mixing. We find that if the $Z^\prime$ boson couples universally to the mirror neutrinos, then the kinetic $Z - Z'$ mixing does not contribute to the active-sterile mixings. Similarly the single Higgs doublet version of the model also does not lead to $\nu-\nu'$ mixing. However, once there are more than one Higgs doublet in each sector, the ``large" $\nu-\nu'$ mixing arises. Crucial to this is the four-scalar mixing of type $(\phi^\dagger\phi)(\phi^{'\dagger}\phi')$  in the Higgs potential as in Eq.~(\ref{mixing-Lagrangian}). The relevant coefficients are estimated to be of order $10^{-3} $.
\end{enumerate}
With explicit formulas and reasonable simplifications we demonstrate in detail how this model can become a natural candidate for the 3+1 scenario. With the simple numerical examples we show that this model can easily fit the present active and sterile neutrino data.

\begin{acknowledgments}
The authors would like to thank  Haipeng An, Yue Zhang and Shmuel Nussinov for helpful discussions.
We also appreciate the valuable comments by Wei Chao.
This work  of R. N. M. is supported by National Science Foundation grant No. PHY-0968854.
The work of Y. Z. is partially supported by the National Natural Science Foundation of China (NSFC) under Grant No. 11105004.
\end{acknowledgments}

\appendix
\section{On the mirror quark masses}
If the mirror electroweak scale is $10^{4}$ times larger than in the visible sector, then one may worry about that the mirror up and down quark masses might be a few times larger than the expected mirror nucleon mass of 5 GeV.
Setting the QCD coupling in the two sectors of ADM to be the same at a ultrahigh energy scale as in~\cite{ADM-An},  we estimate the mirror quark masses, $m_{u'}\simeq 20$ GeV and $m_{d'}\simeq 23$ GeV. With these given values, to obtain a 5 to 10 GeV mirror proton or neutron playing the role of dark matter particle requires additional assumptions: (i) one possibility is to assume a deep gluon potential in the mirror sector to cancel out most part of the quark masses. What we mean by a deep gluon potential is that the gluon contribution to nucleon mass should be large and negative. Since this contribution depends on $\Lambda^\prime_{\rm QCD}$, it may be a possibility. 
(ii) A simpler second possibility is to assume that there are two Higgs doublets in  both the familiar and mirror sectors, giving rise to electroweak symmetry breaking  and asymmetric VEV pattern in the mirror sector in such a way that the ratio of mirror to familiar quark masses does not scale necessarily as $v'_{wk}/v_{wk}= 10^4$.

 We also note other phenomenological facts that may help to alleviate the tension between the predicted dark matter mass in our model and neutrino phenomenology:
\begin{enumerate}
  \item
  The three candidate events recently observed in CDMS II~\cite{CDMS}, together with two other dark matter experiments CoGeNT~\cite{Cogent} and DAMA/LIBRA~\cite{DAMA}, imply a light dark matter with mass around 9 GeV. In light of this, a moderately larger dark matter mass would be more acceptable in our ADM model. Taking into consideration of the fact that $\Omega_{\rm DM}\simeq5\Omega_{{\rm baryonic}}$, the primordial ordinary and mirror leptonic asymmetries are required to be a bit different, which can be easily realized in the ADM model e.g. due to the existence of extra energy release in the mirror sector via heavy neutrino decays, which dilute the mirror baryon density compared to the normal baryon density. 
  \item
  As shown in Sec.~\ref{sec:numrical}, if the matrix elements of $M_D$ are complex and there are moderate cancellations between different parts of the elements in Eq.~(\ref{eqn:Mnu2}), then the scale of $M_{D}$ will be generally a bit larger, e.g. with a $\sim$GeV mass. In this case, as the mirror neutrino masses (and the scale of $M{}_{D}^{\prime}$) are kept unchanged so that the ratio $v_{wk}^{\prime}/v_{wk}$ becomes smaller ($\sim10^{3.5}$),  the mirror quarks could be lighter thereby alleviating any tension.
\end{enumerate}
Taking into consideration all these facts, our model connecting dark matter and the 3+1 neutrino physics is fully compatible with observations and there seems to be no severe flaw or unnaturalness at this point. 

\section{On mirror electron-positron annihilation}

\begin{figure}
  \centering
  \includegraphics{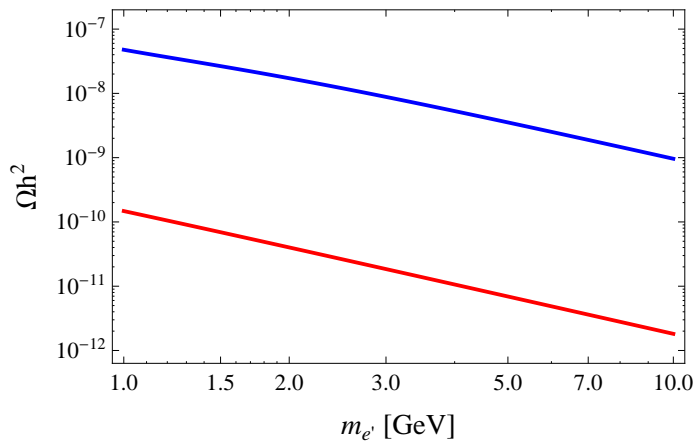}
  \caption{The relic density $\Omega h^{2}$ of the mirror electrons as function of the mirror electron mass $m_{e'}$ in unit of GeV with $m_{\gamma'}=100$ MeV (thick red line) and $m_{\gamma'}=500$ MeV (thick blue line), in the limit of non-relativistic mirror electrons. }
  \label{fig:epa}
\end{figure}

One potential problem in the ADM model is the annihilation of mirror electrons and positrons in the early universe and their subsequent decays. $e^{\prime+}e^{\prime-}$ can annihilate to two mirror photons or the SM particles via $\gamma-\gamma'$ mixing. As the latter channel is suppressed by the small $\gamma-\gamma'$ mixing, we consider only the contribution from the former one. With the representative values of $m_{e'}=5$ GeV and $m_{\gamma'}=100$ MeV ($m_{\gamma'}=500$ MeV) in our model, the annihilation cross section is pretty large, $\sigma\sim10^{-23}\;{\rm cm}^2$ ($\sim10^{-26}\;{\rm cm}^2$). Then we can estimate the relic density of mirror electrons in the universe $\Omega h^2$:
For $m_{e'}=5$ GeV, we have a very small relic density $\Omega h^{2}=7\times10^{-12}$ ($4\times10^{-9}$). Fig.~\ref{fig:epa} tells us explicitly that mirror electron-positron pairs can annihilate fast enough, making a vanishing contribution to the energy budget of the universe.




\begin{thebibliography}{99}

\bibitem{LSND}
  C.~Athanassopoulos {\it et al.}  [LSND Collaboration],
  Phys.\ Rev.\ Lett.\  {\bf 75}, 2650 (1995)
  [nucl-ex/9504002];
  C.~Athanassopoulos {\it et al.}  [LSND Collaboration],
  Phys.\ Rev.\ C {\bf 54}, 2685 (1996)
  [nucl-ex/9605001];
  C.~Athanassopoulos {\it et al.}  [LSND Collaboration],
  Phys.\ Rev.\ Lett.\  {\bf 77}, 3082 (1996)
  [nucl-ex/9605003];
  A.~Aguilar-Arevalo {\it et al.}  [LSND Collaboration],
  Phys.\ Rev.\ D {\bf 64}, 112007 (2001)
  [hep-ex/0104049].

\bibitem{MiniBooNE}
  A.~A.~Aguilar-Arevalo {\it et al.}  [MiniBooNE Collaboration],
  Phys.\ Rev.\ Lett.\  {\bf 102}, 101802 (2009)
  [arXiv:0812.2243 [hep-ex]];
  A.~A.~Aguilar-Arevalo {\it et al.}  [MiniBooNE Collaboration],
  Phys.\ Rev.\ Lett.\  {\bf 105}, 181801 (2010)
  [arXiv:1007.1150 [hep-ex]];
  A.~A.~Aguilar-Arevalo {\it et al.}  [MiniBooNE Collaboration],
  arXiv:1207.4809 [hep-ex];
  A.~A.~Aguilar-Arevalo {\it et al.}  [MiniBooNE Collaboration],
  Phys.\ Rev.\ Lett.\  {\bf 110}, 161801 (2013)
  [arXiv:1303.2588 [hep-ex]].

\bibitem{reactor-anomaly}
  T.~.A.~Mueller {\it et al.},
  Phys.\ Rev.\ C {\bf 83}, 054615 (2011)
  [arXiv:1101.2663 [hep-ex]];
  G.~Mention {\it et al.},
  Phys.\ Rev.\ D {\bf 83}, 073006 (2011)
  [arXiv:1101.2755 [hep-ex]];
  P.~Huber,
  Phys.\ Rev.\ C {\bf 84}, 024617 (2011)
  [arXiv:1106.0687 [hep-ph]].

\bibitem{Gallium-anomaly}
  J.~N.~Abdurashitov {\it et al.},
  Phys.\ Rev.\ C {\bf 73}, 045805 (2006)
  [nucl-ex/0512041];
  C.~Giunti and M.~Laveder,
  Phys.\ Rev.\ C {\bf 83}, 065504 (2011)
  [arXiv:1006.3244 [hep-ph]].

\bibitem{sn-fit-Conrad}
  J.~M.~Conrad, C.~M.~Ignarra, G.~Karagiorgi, M.~H.~Shaevitz and J.~Spitz,
  Adv.\ High Energy Phys.\  {\bf 2013}, 163897 (2013)
  [arXiv:1207.4765 [hep-ex]].
\bibitem{sn-fit-Giunti1}
  M.~Archidiacono, N.~Fornengo, C.~Giunti and A.~Melchiorri,
  Phys.\ Rev.\ D {\bf 86}, 065028 (2012)
  [arXiv:1207.6515 [astro-ph.CO]].
\bibitem{sn-fit-Giunti2}
  C.~Giunti, M.~Laveder, Y.~F.~Li, Q.~Y.~Liu and H.~W.~Long,
  Phys.\ Rev.\ D {\bf 86}, 113014 (2012)
  [arXiv:1210.5715 [hep-ph]];
  M.~Archidiacono, N.~Fornengo, C.~Giunti, S.~Hannestad and A.~Melchiorri,
  arXiv:1302.6720 [astro-ph.CO].
\bibitem{sn-fit-Kopp}
  J.~Kopp, P.~A.~N.~Machado, M.~Maltoni and T.~Schwetz,
  JHEP {\bf 1305}, 050 (2013)
  [arXiv:1303.3011 [hep-ph]].

\bibitem{sn-fit-planck}
  A.~Mirizzi, G.~Mangano, N.~Saviano, E.~Borriello, C.~Giunti, G.~Miele and O.~Pisanti,
  arXiv:1303.5368 [astro-ph.CO].

\bibitem{Planck}
  P.~A.~R.~Ade {\it et al.}  [Planck Collaboration],
  arXiv:1303.5076 [astro-ph.CO].

\bibitem{abaz}
  For references to sterile neutrino models and physics of sterile neutrinos, see
  K. N. Abazajian, M. A. Acero, S. K. Agarwalla, A. A. Aguilar-Arevalo, C. H. Albright, S. Antusch, C. A. Arguelles and A. B. Balantekin et al.,
  arXiv:1204.5379 [hep-ph];
  for references to keV sterile neutrino theory, see A.~Merle,
  arXiv:1302.2625 [hep-ph].
  Much of the discussion in this paper is also applicable to eV sterile neutrinos.


\bibitem{ADM-An}
  H.~An, S.~-L.~Chen, R.~N.~Mohapatra and Y.~Zhang,
  JHEP {\bf 1003}, 124 (2010)
  [arXiv:0911.4463 [hep-ph]].

\bibitem{seesaw}
  P. Minkowski, Phys. Lett. B {\bf 67}, 421 (1977);
  T. Yanagida, {\it Workshop on unified theories and baryon number in the universe}, edited by A. Sawada and A. Sugamoto (KEK, Tsukuba, 1979);
  M. Gell-Mann, P. Ramond and R. Slansky, {\it Supergravity}, edited by P. Van Niewenhuizen and D. Freeman (North Holland, Amsterdam, 1980);
  R. N. Mohapatra and G. Senjanovi\'{c}, Phys. Rev. Lett. {\bf 44}, 912 (1980).

\bibitem{okun}
  For review of the references to mirror models, see
  Z.~Berezhiani,
  Int.\ J.\ Mod.\ Phys.\ A {\bf 19}, 3775 (2004)
  [hep-ph/0312335];
  L.~B.~Okun,
  Phys.\ Usp.\  {\bf 50}, 380 (2007)
  [hep-ph/0606202].

\bibitem{mirrorsterile}
  For other suggestions for identifying the mirror neutrinos as sterile neutrinos see,
  R.~Foot and R.~R.~Volkas,
  Phys.\ Rev.\ D {\bf 52}, 6595 (1995)
  [hep-ph/9505359];
  Z.~G.~Berezhiani and R.~N.~Mohapatra,
  Phys.\ Rev.\ D {\bf 52}, 6607 (1995)
  [hep-ph/9505385];
  K.~S.~Babu and R.~N.~Mohapatra,
  Phys.\ Lett.\ B {\bf 532}, 77 (2002)
  [hep-ph/0201176];
  V.~Berezinsky, M.~Narayan and F.~Vissani,
  Nucl.\ Phys.\ B {\bf 658}, 254 (2003)
  [hep-ph/0210204];
  P.~-H.~Gu,
  arXiv:1303.6545 [hep-ph].

\bibitem{inverse}
  R.~N.~Mohapatra,
  Phys.\ Rev.\ Lett.\  {\bf 56}, 561 (1986);
  R.~N.~Mohapatra and J.~W.~F.~Valle,
  Phys.\ Rev.\ D {\bf 34}, 1642 (1986).

\bibitem{dolgov}
  Z.~G.~Berezhiani, A.~D.~Dolgov and R.~N.~Mohapatra,
  Phys.\ Lett.\ B {\bf 375}, 26 (1996)
  [hep-ph/9511221];
  J.~-W.~Cui, H.~-J.~He, L.~-C.~L\"u and F.~-R.~Yin,
  Phys.\ Rev.\ D {\bf 85}, 096003 (2012)
  [arXiv:1110.6893 [hep-ph]];
  J.~-W.~Cui, H.~-J.~He, L.~-C.~L\"u and F.~-R.~Yin,
  Int.\ J.\ Mod.\ Phys.\ Conf.\ Ser.\  {\bf 10}, 21 (2012)
  [arXiv:1203.0968 [hep-ph]].

\bibitem{MRISM1}
  P.~S.~B.~Dev and A.~Pilaftsis,
  Phys.\ Rev.\ D {\bf 86}, 113001 (2012)
  [arXiv:1209.4051 [hep-ph]];
  P.~S.~B.~Dev and A.~Pilaftsis,
  arXiv:1212.3808 [hep-ph].

\bibitem{Weinberg-d=5}
  S.~Weinberg,
  Phys.\ Rev.\ Lett.\  {\bf 43}, 1566 (1979);
  E.~K.~Akhmedov, Z.~G.~Berezhiani and G.~Senjanovic,
  Phys.\ Rev.\ Lett.\  {\bf 69}, 3013 (1992)
  [hep-ph/9205230].

\bibitem{FNV}
  A.~Denner, H.~Eck, O.~Hahn and J.~Kublbeck,
  Nucl.\ Phys.\ B {\bf 387}, 467 (1992).

\bibitem{FCNC}
  D.~Chang, W.~S.~Hou and W.~-Y.~Keung,
  Phys.\ Rev.\ D {\bf 48}, 217 (1993)
  [hep-ph/9302267].

\bibitem{MEG}
  J.~Adam {\it et al.}  [MEG Collaboration],
  arXiv:1303.0754 [hep-ex].

\bibitem{nu-fit}
  D.~V.~Forero, M.~Tortola and J.~W.~F.~Valle,
  Phys.\ Rev.\ D {\bf 86}, 073012 (2012)
  [arXiv:1205.4018 [hep-ph]].

\bibitem{sn-fit-theta34}
  T. Schwetz, talk at the conference Sterile Neutrinos at Crossroads, Virginia Tech, USA, 2011.

\bibitem{nuless}
  For recent reviews, see W.~Rodejohann,
  J.\ Phys.\ G {\bf 39}, 124008 (2012)
  [arXiv:1206.2560 [hep-ph]]; J.~D.~Vergados, H.~Ejiri and F.~Simkovic,
  Rept.\ Prog.\ Phys.\  {\bf 75}, 106301 (2012)
  [arXiv:1205.0649 [hep-ph]].


\bibitem{CDMS}
  R.~Agnese {\it et al.}  [CDMS Collaboration],
  [arXiv:1304.4279 [hep-ex]].

\bibitem{Cogent}
  C.~E.~Aalseth {\it et al.}  [CoGeNT Collaboration],
  Phys.\ Rev.\ Lett.\  {\bf 106}, 131301 (2011)
  [arXiv:1002.4703 [astro-ph.CO]];
  C.~E.~Aalseth, P.~S.~Barbeau, J.~Colaresi, J.~I.~Collar, J.~Diaz Leon, J.~E.~Fast, N.~Fields and T.~W.~Hossbach {\it et al.},
  Phys.\ Rev.\ Lett.\  {\bf 107}, 141301 (2011)
  [arXiv:1106.0650 [astro-ph.CO]].

\bibitem{DAMA}
  R.~Bernabei {\it et al.}  [DAMA and LIBRA Collaborations],
  Eur.\ Phys.\ J.\ C {\bf 67}, 39 (2010)
  [arXiv:1002.1028 [astro-ph.GA]].


\end{thebibliography}
\end{document}